\documentclass[twocolumn,showpacs,preprintnumbers,amsmath,amssymb]{revtex4}

\usepackage{graphicx}
\usepackage{dcolumn}
\usepackage{bm}

\newcommand{\be}{\begin{equation}}
\newcommand{\ee}{\end{equation}}

\begin{document}

\title{ Melting temperature of screened Wigner crystal on 
helium films by molecular dynamics}

\author{J. A. R. da Cunha and Ladir C\^andido }
\affiliation{Instituto de F\'isica, Universidade Federal de Goi\'as
Campus Samambaia, 74001-970 Goi\^ania, GO, Brazil}

 \vspace*{3mm}

\begin{abstract}
Using molecular dynamics (MD) simulation, we have calculated the
melting temperature of two-dimensional electron systems on
$ 240$\AA-$ 500$\AA \ helium films supported by substrates
of dielectric constants  $ \epsilon_{s}=2.2-11.9$ at areal
densities $n$ varying from $ 3\times 10^{9}$ cm$^{-2}$ to
$ 1.3\times 10^{10}$ cm$^{-2}$. Our  results are in good agreement
with the available theoretical and experimental results.

\end{abstract}


\maketitle

At suficiently low densities and temperatures, an electron gas
is expected to undergo a phase transition to a lattice (because of
the domination of the Coulomb interaction energy over the kinetic
energy) which has received the name Wigner crystal\cite{wigner}. The
two-dimensional (2d) Wigner crystal is well established and experimentally
it was first observed on liquid helium surface\cite{grimes} and more recently
in semiconductors structures like MOSFET's and heterojunctions\cite{yoon}. These
systems can be used for testing several theoretical predictions in many-body
theory, such as phase-transitions of the electron system, metal-insulator
transition and now electrons on helium surface are being proposed as
a set of strongly interacting quantum bits for quantum computers\cite{platzman}.
Electrons on the surface of bulk helium form a crystal at a temperature
$T_{m}=2e^2(\pi n)^{1/2}/(\epsilon_{He}+1)\Gamma_{m}$,
which is much higher than the Fermi temperature, $T_{F}=\pi n\hbar^{2}/m$
in a density range of $10^{5}-10^{9}$ cm$^{-2}$ (where $n$ is the
electron areal density, $\epsilon_{He}$ is the dieletric
constant of helium and $\Gamma_{m}$ is the plasma parameter in the melting
temperature defined as the ratio of potential to kinetic energy).
Therefore, such electrons in this regime obey the classical Boltzmann
statistics. Experimentally the liquid to solid transition in the bulk
takes place for a value of the coupling constant
$\Gamma_{m}=137\pm15$\cite{grimes} and computer simulations of
Kalia {\it et al.}\cite{kalia} showed an agreement with the experimental
measurements indicating a first-order melting in $\Gamma_{m}=118-130$.

Superficial electrons on liquid helium films form also a very
interesting system to study the many-body properties of 2d screened
systems. In this case the screening is provided by the image
charges in the substrate beneath the film. The screening effect
can drastically change the electron-electron interacting potential
going from $1/r$ to $1/r^{3}$ through varying external
parameters such as film thickness and dielectric constant of the
substrate. Peeters \cite{peeters} using a phenomenological approach got
a reduction in the phase diagram of this electron system comparing with
the bulk case. Saitoh\cite{saitoh} obtained the melting transition in this
system using an analytical approximation to the angular frequency of the
transverse Wigner phonon combined with the Kosterlitz-Thoules melting
criterion. His result is in agreement with the experiment by Jiang
{\it et al.}\cite{jiang}. C\^andido {\it et al.}\cite{cand} studied
the thermodynamical, structural and dynamical properties of this
two-dimensional electron system by computer simulation.
Experimmentally, the melting temperature of the Wigner crystal on
thin helium films adsorbed on dielectric substrates was measured
by Jiang {\it et al.}\cite{jiang} through electron mobility
and by Mistura {\it et al.}\cite{mist} using microwave cavity
technique.

In this paper, we present an accurate MD calculation for the melting
temperature for an electron system over a helium film adsorbed
on a dielectric substrate. In Fig. 1 we show schematically the
geometrical arrangement of the system considered. The obtained results
are directly compared with the available experimental data,
of Mistura {\it et al.}\cite{mist} and Jiang {\it et al.}\cite{jiang},
and the theoretical results of Peeters\cite{peeters} and Saitoh\cite{saitoh}.

We consider a two-dimensional system of electrons on a helium
film of thickness $d$ adsorbed on a substrate of dielectric
constant $\epsilon_{s}$, interacting through a screening Coulomb
potential\cite{smythe}. The electron system is immersed in a rigid,
uniform, positively charged background to make a neutral
charged system. The Hamiltonian for such a system is given by

\begin{equation}
 H = \frac{1}{2}m\sum_{i}{v}^{2}_{i} +
 \sum_{i > j}e^2\left[\frac{1}{r}_{ij} -
\frac{\delta}{\sqrt{r_{ij}^{2}+(2d)^{2}}}\right] + U_{b},
\end{equation}
where $\delta=(\epsilon_{s}-1)/(\epsilon_{s}+1)$ with the dielectric constant
of helium approximated by 1 $(\epsilon_{He}=1.057)$ and $ U_{b}$ is the
interaction of electrons with the uniform positively charged background.

\begin{figure}
\includegraphics[width=8cm,height=4cm]{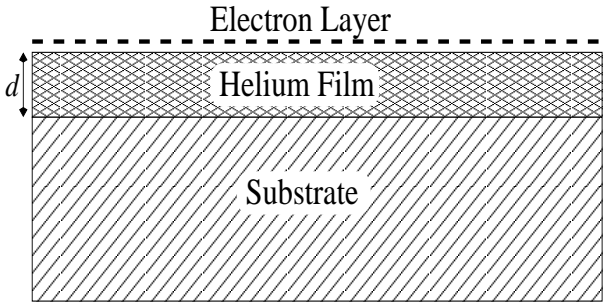}
\caption{
Schematic view of the electron system.}
\end{figure}

In this work most of the molecular dynamics calculations were
performed on a system of $100$ electrons with a few runs of $484$
and $784$ electrons to study size effect. The finite size effect
is investigated by changing the system size and the thermodynamical
behavior in an infinite system is derived from their extrapolation.
The initial position of the electrons is a triangular lattice which is
accomodated in a rectangular box with periodic boundary condictions to
eliminate the surface effects. Because of the long range nature of the
electron-electron and electron-background interacting potential we are
employing Ewald summation which splits the potential into a
long-range and a short-range part. The long-range part is handled
in the k-space, while the short range part in the real space.
We have used the fifth-order predictor-corrector algoritm to integrate
the Newton's equation of motion with the MD time step varying from
$10^{-12}$ to $10^{-15}$ sec, since it has some scale dependence on
the electron densities. The optimum time step leads to a conservation of
the total energy of 1 part in $10^{4}$ after several thousand time step
runs. The time average of the physical quantities were obtained over
$120$ 000 time steps after the system has reached equilibrium.

In Fig. 2 we present the results for the total energy per electron
versus temperature to illustrate the general feature of the melting transition
in this system. The solid squares in the figure represent the results for the
electron liquid which has been monotonically cooled from a higher temperature.
The open circles are the results for the electron solid which has
been monotonically heated from a lattice at very low temperatures.
It means that our simulations were performed in cascade, i.e., an equilibrated
configuration obtained for a given higher (lower) temperature was used as an
input to reach another configuration at lower (higher) temperature.
As one can see the electron system shows hysteresis and latent
heat on melting, which characterize a first order transition as other 2d
classical systems. The melting temperature range is  $1.83$ K $< T <$ $2.05$ K 
defined from the vertical dashed lines in Fig. 2 representing the hysteresis
region. Thus, we would define the melting temperature, $T_{m}$,
as exactly the mean point in the temperature width of the hysteresis
$\Delta T$, i.e, $T_{m}=(1.94\pm 0.11)$ K with the error bar giving by
half the temperature width of the hysteresis. The value of the latent heat
per particle and the change in the entropy on melting are found to be
$0.40$ K and  $0.21$ k$_{B}$, respectively. We also find our MD results
for the melting temperature is in agreement with those of Kalia
{\it et al.}\cite{kalia} for the bulk limit.

\begin{figure}
\includegraphics[width=8cm,height=6cm]{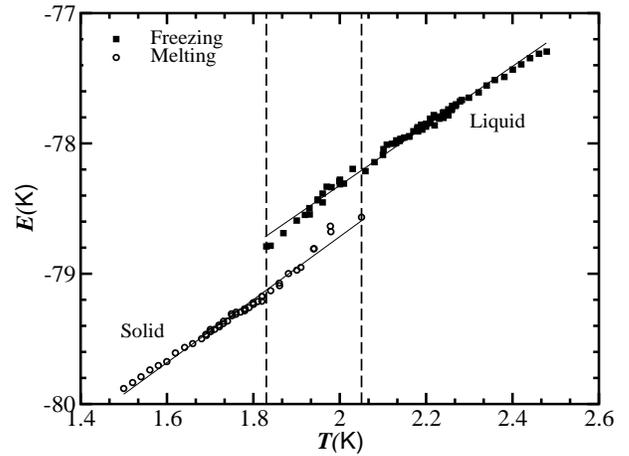}
\caption{Total energy per electron as a function of temperature for
a system of $N=100$ electrons on a helium film supported by a
glass substrate, $\epsilon_{s}=7.3$, film thickness $d=240$ \AA \ and
density $n=1.3$x$10^{10}$ cm$^{-2}$ .}
\end{figure}

Fig. 3 shows size dependences of the transition temperature
$T_{m}$ for different electron densities. The error
bars on $T_{m}$ indicate the hysteresis width. When electron
number becomes larger, $T_{m}$ decreases because the periodic boundary
conditions favors the solid phase. The transition temperature, however,
seems to follow a linear decrease as a function of $1/N$. Therefore the
melting point in the thermodynamic limit can be obtained definitely
by extrapolating the finite size data.

\begin{figure}
\includegraphics[width=8cm,height=6cm]{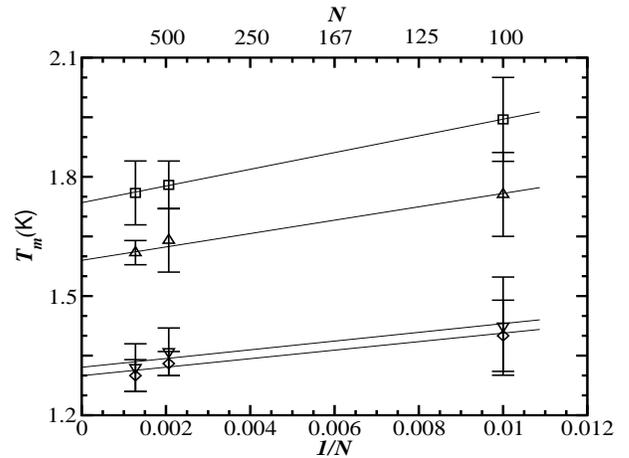}
\caption{
Size dependence of the melting temperature for electrons on helium films
above a substrate with dielectric constant $\epsilon_s=7.3$ at four
different densities: $n=1.3\times 10^{10}$ cm$^{-2}$ and
$d=240$ \AA (squares); $n=1.0\times 10^{10}$ cm$^{-2}$ and
$d=285$ \AA (triangles-up); $n=0.9\times 10^{10}$ cm$^{-2}$
and $d=260$ \AA (triangles-down); and $n=0.75\times 10^{10}$ cm$^{-2}$
and $d=305$ \AA (diamonds). The lines are linear fit.}
\end{figure}

The extrapolated melting temperature is exhibited in Fig. 4 as a function of
the electron density (top), film thickness (middle) and dielectric constant of
the substrate (bottom). We roughly estimated the error bar on the
experimental values for the melting temperature to indicate the
uncertainty of about $15\% - 20\%$ on the experimental measurement of the
electron density. As is shown in Fig. 4, our results are in good agreement
with those obtained experimentally in Ref.\cite{mist}.
The top panel shows that the melting temperature increases with
increasing the electron density (for fixed $\epsilon_{s}$ and $d$)
due to the the fact that the screening becomes weaker and, consequently,
the electron-electron interaction is enhanced. The middle panel also shows a
shift in the melting transition to higher temperature with increasing the
film thickness (for fixed n and $\epsilon_{s}$), which is a
consequence of decreasing in the screening resulting in a stronger
electron-electron interaction.
In the bottom panel, the melting temperature decreases as the dielectric
constant of the substrate increases at fixed $d$ and $n$.
A larger dielectric constant of the substrate leads to a more polarizable
system with stronger screening. As a consequence, the melting temperature goes
down.

\begin{figure}
\includegraphics[width=6.cm,height=12cm]{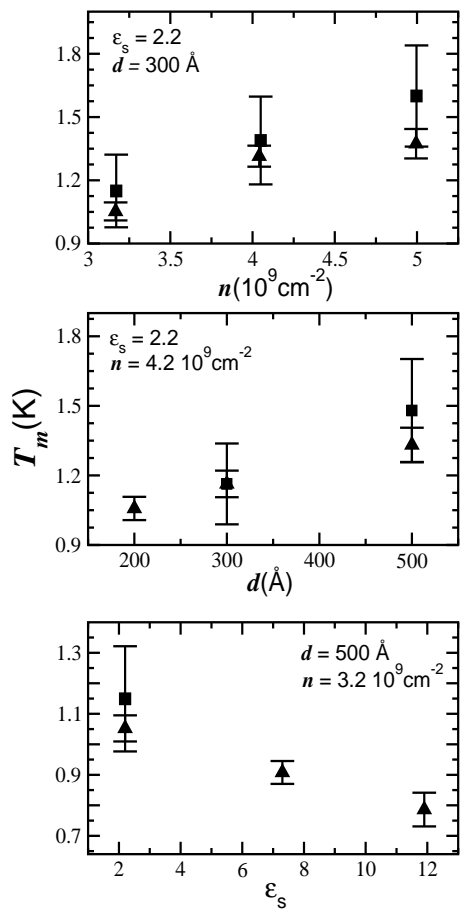}
\caption{
The melting temperature as a function of the electron density (top),
film thickness (middle) and dielectric constant of the substrate (bottom). The
experimental results from Ref.\cite{mist} are given by solid squares and our MD
simulation results are indicated by solid triangles.}
\end{figure}

In table I, we show a comparison of our MD simulation results of the melting
temperature with the available theoretical and experimental results
for electron systems on a thin helium film surface. For densities below
$1.0 \times 10^{10}$ cm$^{-2}$ one can see that our MD calculations are
in agreement with both Jiang {\it et al.} \cite{jiang} and 
Mistura {\it et al.} \cite{mist}
experimental measurements. They are also in agreement with Saitoh's
theoretical results\cite{saitoh} in the range of the densities studied.
However, we got some discrepancies with Peeters's results\cite{peeters}
that can be justified, as pointed out by Saitoh\cite{saitoh}, as being a
possible double counting on $T_m$. For densities larger than
$1.0\times 10^{10}$ cm$^{-2}$, our MD simulation melting temperatures
are higher than the experimental ones, though the differences are almost
within the uncertainty of the experimental results. A possible explanation
for this discrepancy is that the quantum effect can be important at
such densities. Besides, we note that the change in entropy on melting
decreases as the density or the dielectric constant of the substrate
(the film thichness) increases (decreases). This might imply that transition
becomes continuous at high densities.

\begin{table*}
\caption{Data of melting temperature $T_{m}$ for different thicknesses
$d$, dielectric constant of the substrates $\epsilon_{s}$ and electron
densities $n$. The experimental uncertainty in the absolute value of the
electron density is about $15\% - 20\%$ in Refs.\cite{jiang} and
\cite{mist}. Quantities in ( ) are the estimated error on $T_{m}$ defined
as half of temperature width of the hysteresis, in the last decimal place.
}
\begin{tabular}{|c|c|c|ccccc|}
\hline
\hline
 $ \epsilon_{s} $&  $d$( \AA )& $n$ ($10^{10}$cm$^{-2}$) & $  $& $  $
 &   $T_{m}$(K)   & $ $ & $ $\\
 & & & $$ Mistura et al.\cite{mist}$$& $$ Jiang {\it et al.}\cite{jiang} $$
 & $$ Peeters\cite{peeters}$$ & $$ Saitoh\cite{saitoh}  $$ & $$ MD - this work $$\\
 \hline
 2.2  & 300 & 0.42 & 1.16 &  --  &  --  & 1.08 & 1.16(6)\\
 2.2  & 300 & 0.53 & 1.43 &  --  &  --  & 1.21 & 1.26(7)\\
 2.2  & 300 & 0.57 & 1.50 &  --  &  --  & 1.26 & 1.33(6)\\
 2.2  & 500 & 0.32 & 1.15 &  --  &  --  & 1.05 & 1.05(4)\\
 2.2  & 500 & 0.40 & 1.39 &  --  &  --  & 1.17 & 1.31(5)\\
 2.2  & 500 & 0.50 & 1.60 &  --  &  --  & 1.29 & 1.37(7)\\
 3.9  & 237 & 1.05 & 1.16 &  --  &  --  &  --  & 1.64(9)\\
 7.3  & 305 & 0.75 & --   & 1.23 & 0.85 & 1.24 & 1.29(5)\\
 7.3  & 260 & 0.90 & --   & 1.28 & 0.89 & 1.31 & 1.32(8)\\
 7.3  & 285 & 1.00 & --   & 1.32 & 1.02 & 1.54 & 1.59(7)\\
 7.3  & 240 & 1.30 & --   & 1.38 & 1.14 & 1.62 & 1.73(8)\\
\hline
\end{tabular}
\end{table*}

In conclusion, we have shown that the MD is able to reproduce
the experimental measurements of the melting temperature in the
two-dimensional electrons on thin liquid He films. Our results are
in good agreement with those obtained by Mistura
{\it et al.}\cite{mist} and Jiang {\it et al.}\cite{jiang}. These
results in the classical regime ({\it i.e.} $ n \leq 1.0\times 10^{10}$
cm$^{-2}$) should be useful to the experimental and
theoretical investigation of the melting transition in this system.
For larger densities ( $  n >  1.0\times 10^{10}$ cm$^{-2}$ and
$\epsilon_{s} > 2.2$), the results might be beyond the applicability
of the present method.

\begin{acknowledgments}
This research was partially sponsored by Conselho Nacional de
Desenvolvimento Cient\'{i}fico e Tecnol\'{o}gico (CNPq) and Funda\c c\~ao
Nacional de Apoio a Pesquisa (FUNAPE-UFG). J.A.R.C is
supported by Funda\c c\~ao Coordenac\~ao de Aperfei\c coamento de Pessoal
de N\'{i}vel Superior (CAPES). We are grateful to G.-Q. Hai for useful
discussions.
\end{acknowledgments}

\end{document}